\documentclass[a4paper,12pt,oneside]{article}
\usepackage[latin1]{inputenc}
\usepackage[T1]{fontenc}
\usepackage[english]{babel}
\usepackage[mathscr]{euscript}
\usepackage{amsfonts,amssymb,amsthm,latexsym,amsmath,exscale}
\begin{document}
\title{Simple physical applications of a groupoid structure}
\author{Giuseppe Iurato}
\date{}\maketitle\begin{abstract}Motivated by Quantum Mechanics considerations,
we expose some cross product constructions on a groupoid structure. Furthermore, critical remarks are made on
some basic formal aspects of the Hopf algebra structure.\end{abstract}$$\normalsize\bf 1.\ Introduction$$

In [11] was recovered the notion of \it E-groupoid \rm [\it EBB-groupoid\rm], whose groupoid algebra led to the
so-called \it E-groupoid algebra \rm[\it EBB-groupoid algebra\rm]. Following a suggestion of A. Connes (see [6,
§ I.1]), some elementary algebraic structures of Matrix Quantum Mechanics can arise from certain representations
of such groupoids and related group algebras. For instance, the map $(i,j)\rightarrow\nu_{ij}$ of [11, § 5],
provides the kinematical time evolution of the observables $\mathfrak{q}$ and $\mathfrak{p}$, given by the
Hermitian matrices [11, § 5, $(\maltese)$], that, by means of the Kuhn-Thomas relation, satisfies the celebrated
Heisenberg canonical commutation relations (see [23, § 14.4] and [3, § 2.3]), whereas the HBJ EBB-groupoid
algebra constructed in [11, § 6], is the algebra of physical observables according to W. Heisenberg.\\

The structure of groupoid has many considerable applications both in pure and applied mathematics (see [4, 25]):
here, let us consider the following, particular example, drew from Quantum Field Theory (QFT).\\

\footnotesize $\bullet$ \it A toy model of QFT. \rm In Renormalization of Quantum Field Theory (see [26] and
references therein), a Feynman graph $\Gamma$ may be analytically represented as sum of iterated divergent
integrals defined as follows
$$\Gamma^1(t)=\int_t^{\infty}\frac{dp_1}{p_1^{1+\varepsilon}},\ \ \ \Gamma^n(t)=\int_t^{\infty}\frac{dp_1}
{p_1^{1+\varepsilon}}\int_{p_1}^{\infty}\frac{dp_2}{p_2^{1+\varepsilon}}\ ...\int_{p_{n-1}}^{\infty}
\frac{dp_n}{p_n^{1+\varepsilon}}\ \ \ n\geq 2$$for $\varepsilon\in\mathbb{R}^+$; such integrals diverges
logarithmically for $\varepsilon\rightarrow 0^+$.\\It can be proved as these iterated integrals form a Hopf
algebra of rooted trees (see [7, 26]), say $\mathcal{H}_R$, and called the \it Connes-Kreimer Hopf algebra \rm
(see [7]). The renormalization of these integrals requires a regularization, for instance through a linear
multiplicative functional $\phi_a$ (bare Green function, defining a \it Feynman rule\rm) defined on them, which
represents a certain way of evaluation of the Feynman graphs, at the energy scale $a$, of the
type$$\phi_a\Big(\prod_{i\in I}\Gamma^i(t)\Big) =\prod_{i\in I}\Gamma^i(a),$$where $I$ denotes an arbitrary
finite ordered subset of $\mathbb{N}$, and $\Gamma_a=\Gamma^0(a)$ are the normalized coupling constants at the
energy scale (or renormalization point) $a$; if $\Gamma$ is a Feynman graph, then $\phi_a(\Gamma)$ is the
corresponding regularized Feynman amplitude, according to the renormalization scheme parametrized by $a$.\\So,
every Feynman rule is a character $\phi_a:\mathcal{H}_R\rightarrow\mathbb{C}$ of the Hopf algebra
$\mathcal{H}_R$, and their set is a (renormalization) group $\mathcal{G}_R$ under the group law given by the
usual convolution law $\hat{\ast}$ of $\mathcal{H}_R$. Thus, the coalgebra structure of $\mathcal{H}_R$ endows
$\mathcal{G}_R$ with a well-defined group structure.\\ If $S$ denotes the antipode of $\mathcal{H}_R$, then let
us consider the following \it deformed antipode $S_a\doteq\phi_a\circ S$\rm; in [26], it is considered a
particular modification of the usual antipode axiom $S\hat{\ast}\mbox{\rm id}=\mbox{\rm
id}\hat{\ast}S=\eta\circ\varepsilon$ (see [11, § 8]), precisely$$\varepsilon_{a,b}=S_a\hat{\ast}\mbox{\rm
id}_b\doteq(\phi_a\circ S)\hat{\ast}\phi_b=m\circ(S_a\otimes\phi_b)\circ\Delta\quad\mbox{\rm (renormalized Green
functions)}.$$Hence, in [26, § 5], it is proved to be true the following pair groupoid law
$\varepsilon_{a,b}\hat{\ast}\varepsilon_{b,c}=\varepsilon_{a,c}$, deduced from the Hopf algebra properties of
$\mathcal{H}_R$. Moreover, if we consider the renormalized quantities
$\varepsilon_{a,b}(\Gamma^n(t))=\Gamma_{a,b}^n$, then we have
$$\Gamma_{a,b}^1=\int_b^a\frac{dp}{p^{1+\varepsilon}},\ \
\Gamma_{a,b}^2=\int_b^a\frac{dp_1}{p_1^{1+\varepsilon}}\int_{p_1}^a\frac{dp_2}{p_2^{1+\varepsilon}},\ \ ....\
,$$ with every $\Gamma_{a,b}^n$ finite for $\varepsilon\rightarrow 0^+,$ and zero for $a=b$. $\varepsilon_{a,b}$
is said to be a \it renormalized character \rm of $\mathcal{H}_R$ at the energy scales $a,b$. The correspondence
(renormalization schemes) $\phi_b,\phi_b\rightarrow\varepsilon_{a,b}$ is what renormalization typically
achieves.\\Finally, from the relation $\varepsilon_{a,b}\hat{\ast}\varepsilon_{b,c}=\varepsilon_{a,c}$ and the
coproduct rule of $\mathcal{H}_R$, it is possible to obtain the following relation
$$\Gamma^i_{a,c}=\Gamma^i_{a,b}+\Gamma_{b,c}^i+\sum_{j=1}^{i-1}\Gamma_{a,b}^j\Gamma_{b,c}^{i-j}\ \ \ \ i\geq
2,$$that is a generalization of the so-called \it Chen's Lemma \rm (see [5, 13]); this relation describes what
happens if we change the renormalization point.\\If $\Gamma\in\mathcal{H}_R$ is a Feynman graph, then we have
the following asymptotic expansion $\Gamma=\Gamma^0+\Gamma^1+\Gamma^2+\Gamma^3+...$; in general, such a series
may be divergent, and, in this case, it can be renormalized to an finite, but undetermined, value. We have that
$\varepsilon_{a,b}(\Gamma)=\Gamma_{a,b}$ is the result of the regularization of $\Gamma$ at the energy scales
$a,b$, and, respect to the scale change $\phi_a\rightarrow\phi_b$ (which allows us to renormalizes in a
non-trivial manner), we have the following rule for the shift of the normalized coupling constants
$\Gamma_b=\Gamma_a+\sum_{i\in\mathbb{N}}\Gamma_{a,b}^i$, in dependence of the running coupling constants
$\Gamma_{a,b}^i,\ \ i\geq 1$.\\In short, the comparison among different renormalization schemes (via the
variation of the renormalization point) is regulated by the fundamental pair groupoid law
$\varepsilon_{a,b}\hat{\ast}\varepsilon_{b,c}=\varepsilon_{a,c}$. Furthermore, we point out as this groupoid
combination law, connected with a variation of the renormalization points, leads us to further formal properties
of renormalization\footnote{Moreover, it should be interesting to go into the question related to possible,
further roles that the groupoid structures may play in Renormalization. See, also, the conclusions of § 7 of the
present paper.}, as, for instance, the cohomological ones or the Callan-Symanzik type equations.\\

\normalsize In [11, § 5], we have defined a specific EBB-groupoid, called the \it Heisenberg-Born-Jordan
EBB-groupoid \rm (or \it HBJ EBB-groupoid\rm), whose group algebra, said \it HBJ EBB-groupoid algebra, \rm may
be endowed with a (albeit trivial) Hopf algebra structure, obtaining the so-called \it HBJ EBB-Hopf algebra \rm
(or \it HBJ EBBH-algebra\rm), that is a first, possible example of generalization of the structure of Hopf
algebra: indeed, it is a particular weak Hopf algebra, or quantum groupoid (see [19, § 2.5], [20, § 2.1.4] and
[24, § 2.2]), in the finite-dimensional case.\\ We remember that group algebras were basic examples of Hopf
algebras\footnote{See, for instance, the formalization of the quantum mechanics motivations adduced by V. G.
Drinfeld in [9, § 1].}, so that groupoid algebras may be considered as basic examples of a class of structures
generalizing the ordinary Hopf algebra structure; this class contains the so-called weak Hopf algebras, the Lu's
and Xu's Hopf algebroids, and so on (see [1, 2, 14, 27]).\\

In this paper, starting from the E-groupoid structure exposed in [11], we want to introduce another, possible
generalization of the ordinary Hopf algebra structure, following the notions of commutative Hopf algebroid (see
[22]) and of quantum semigroup (see [9, § 1, p. 800]).\\

Moreover, at the end of [11, § 8], it has been mentioned both the triviality of the Hopf algebra structure there
introduced (on the HBJ EBB-algebra), and some non-trivial duality questions related to the (possible) not finite
generation of the HBJ EBB-algebra.\\

At the § 5. of the present paper\footnote{That must be considered as a necessary continuation of [1].}, we'll
try to settle these questions by means of some fundamental works of S. Majid.\\

Indeed, in [16] and [17], Majid has constructed non-trivial examples of non-commutative and non-cocommutative
Hopf algebras (hence, non-trivial examples of quantum groups), via his notion of bicrossproduct.\\This type of
structures involves group algebras and their duals; furthermore, these structures has an interesting physical
meaning, since they are an algebraic representation of some quantum mechanics problems (see also [18, Chap.
6]).\\

Finally, we'll recall some other cross product constructions, among which the group Weyl algebra and the
(Drinfeld) quantum double, that provides further, non-trivial examples of a quantum group having a really
physical meaning.\\ Moreover, if we consider such structures applied to the EBJ EBBH-algebra (of [11, § 8]),
then it is possible to get structures that represents an algebraic formalization of some possible quantum
mechanics problems on a groupoid, in such a way to obtain new examples of elementary structures of a Quantum
Mechanics on groupoids.$$\bf 2.\ The\ Notion\ of\ \mbox{\bf E-semigroupoid}$$

For the notions of E-groupoid and EBB-groupoid, with relative notations, we refer to [11, § 1].\\\\An \it
E-semigroupoid \rm is an algebraic system of the type $(G,G^{(0)},G^{(1)},r,s,i,\star)$, where
$G,G^{(0)},G^{(1)}$ are non-void sets such that $G^{(0)},G^{(1)}\subseteq G$, $r,s:G\rightarrow G^{(0)}$,
$i:G^{(1)}\rightarrow G^{(1)}$ and $G^{(2)}=\{(g_1,g_2)\in G\times G;s(g_1)=r(g_2)\}$, satisfying the following
conditions\footnote{Whenever the relative $\star$-products are well-defined.}:\begin{itemize}\item$_1$
$s(g_1\star g_2)=s(g_2), r(g_1\star g_2)=r(g_1),\ \ \forall (g_1,g_2)\in G^{(2)}$;\item$_2$ $s(g)=r(g)=g,\ \
\forall g\in G^{(0)}$;\item$_3$ $g\star\alpha(s(g))=\alpha(r(g))\star g=g,\ \ \forall g\in G$;\item$_4$
$(g_1\star g_2)\star g_3=g_1\star(g_2\star g_3),\ \ \forall g_1,g_2,g_3\in G$;\item$_5$ $\forall g\in G^{(1)},
\exists g^{-1}\in G^{(1)}: g\star g^{-1}=\alpha(r(g)),g^{-1}\star g=\alpha(s(g)),$\end{itemize}being
$\alpha:G^{(0)}\hookrightarrow G$ the immersion of $G^{(0)}$ into $G$, and $i:g\rightarrow g^{-1}$. The maps
$r,s$ are called, respectively, \it range \rm and \it source, \rm $G$ is the \it support, \rm $G^{(0)}$ is the
\it set of units, \rm and $G^{(1)}$ is the \it set of inverses, \rm of the given E-semigroupoid.\\For
simplicity, we write $r(g),s(g)$ instead of $\alpha(r(g)),\alpha(s(g))$.\\

We obtain an E-groupoid when $G^{(1)}=G$ (see [11, § 1]), whereas we obtain a monoid when $G^{(0)}=\{e\}$.
Moreover, if an E-semigroupoid also verify the condition of [11, § 1, $\bullet_6$], then we have an \it
EBB-semigroupoid. \rm
$$\bf 3.\ The\ Notion\ of\ Linear\ \mbox{\bf $\mathbb{K}$-algebroid}$$

A linear algebra (over a commutative scalar field $\mathbb{K}$) is an algebraic system of the type
$(V_{\mathbb{K}},+,\cdot,m,\eta)$, where $(V_{\mathbb{K}},+,\cdot)$ is a $\mathbb{K}$-linear space and $(V,+,m)$
is a unital ring, satisfying certain
compatibility conditions; in particular, $(V,m)$ is a unital semigroup (that is, a monoid).\\

Following, in part, [21] (where it is introduced the notion of vector groupoid), if it is given an
E-semigroupoid $(G,G^{(0)},G^{(1)},r,s,i,\star)$ such that\begin{description}\item \it i)
$G_{\mathbb{K}}=(G,+,\cdot)$ \rm is a $\mathbb{K}$-linear space, and $G^{(0)},G^{(1)}$ are its linear
subspaces;\item \it ii) $r,s$ \rm and $i$, are linear maps;\item \it iii) \rm $g_1\star(\lambda g_2+\mu
g_3-s(g_1))=\lambda(g_1\star g_2)+\mu (g_1\star g_3)-g_1$, $(\lambda g_1+\mu g_2-r(g_3))\star
g_3=\lambda(g_1\star g_3)+\mu(g_2\star g_3)-g_3,$ for every $g_1,g_2,g_3\in G$ and $\lambda,\mu\in\mathbb{K}$
for which there exists the relative $\star$-products,\end{description}then we say that
$(G_{\mathbb{K}},G^{(0)},G^{(1)},r,s,i,\star)$ is a \it linear $\mathbb{K}$-algebroid. \rm
$$\bf 4.\ The\ Notion\ of\ \mbox{\bf E-Hopf Algebroid}$$

Let $\mathfrak{G}_{\mathbb{K}}=(G_{\mathbb{K}},G^{(0)},G^{(1)},r,s,i,\star)$ be a linear $\mathbb{K}$-algebroid.
If we set$$G^{(2)}=\{g_1\otimes_{\star} g_2\in G\times G;s(g_1)=r(g_2)\}\doteq G\otimes_{\star}G,$$
$$m_{\star}(g_1\otimes_{\star}g_2)=g_1\star g_2,$$then, more specifically, with $(\mathfrak{G}_{\mathbb{K}},m_{\star},
\{\eta_r^{(e)}\}_{e\in G^{(0)}}, \{\eta_s^{(e)}\}_{e\in G^{(0)}})$, we'll denote such a linear
$\mathbb{K}$-algebroid where, for each $e\in G^{(0)}$, we put $\eta_r^{(e)},\eta_s^{(e)}:\mathbb{K}\rightarrow
G$ in such a way that $\eta_r^{(e)}(k)=\{e\}$ and $\eta_s^{(e)}(k)=\{e\},\ \ \forall k\in\mathbb{K}$. Hence, the
unitary and associativity properties $\bullet_3$ and $\bullet_4$, of the given linear $\mathbb{K}$-algebroid,
are as follows\footnote{Henceforth, every partial $\star$-operation (as $\otimes_{\star}$, and so on) that we
consider, it is assumed to be defined.}
\begin{description}\item 1. $m_{\star}\circ(\eta_r^{(e)}\otimes_{\star}\mbox{id})=m_{\star}\circ(\mbox{id}
\otimes_{\star}\eta_s^{(e')}),\quad\forall e,e'\in G^{(0)},$\item 2. $m_{\star}\circ(\eta_r^{(e)}\otimes_{\star}
m_{\star})= m_{\star}\circ(m_{\star}\otimes_{\star}\eta_s^{(e')}),\quad\forall e,e'\in
G^{(0)},$\end{description}where id is the identity of $G$.\\\\Let us introduce, now, a cosemigroupoid structure
as follows.\\We define a partial comultiplication by a map $\Delta_{\star}:G\rightarrow G\otimes_{\star}G$, in
such a way that, when the following condition holds:\begin{description} \item 3.
$(\mbox{id}\otimes_{\star}\Delta_{\star})\circ\Delta_{\star}=(\Delta_{\star}\otimes_{\star}\mbox{id})\circ
\Delta_{\star}$,\end{description}then we say that $(\mathfrak{G}_{\mathbb{K}},\Delta_{\star})$ is a \it
cosemigroupoid.\\ \rm If we require to subsist suitable homomorphism conditions for the maps
$m_{\star},\Delta_{\star}, \{\eta_r^{(e)}\}_{e\in G^{(0)}}, \{\eta_s^{(e)}\}_{e\in G^{(0)}}$, then we may to
establish a certain \it quantum semigroupoid \rm structure (in analogy to the quantum semigroup structure - see
[9, § 1, p. 800]) on $(\mathfrak{G}_{\mathbb{K}},m_{\star}, \{\eta_r^{(e)}\}_{e\in G^{(0)}},
\{\eta_s^{(e)}\}_{e\in G^{(0)}})$.\\

Following, in part, the notion of commutative Hopf algebroid given in [22, Appendix 1], if we define certain
counits by maps $\varepsilon_r^{(e)},\varepsilon_s^{(e)}:G\rightarrow\mathbb{K}$ for each $e\in G^{(0)}$, then
we may to require that further counit properties holds, chosen among the following\begin{description}\item
$4_1.\ \
(\mbox{id}\otimes_{\star}\varepsilon_r^{(e)})\circ\Delta_{\star}=(\varepsilon_r^{(e')}\otimes_{\star}\mbox{id})
\circ\Delta_{\star}=\mbox{id},\quad\forall e,e'\in G^{(0)}$,\item $4_2.\ \
(\mbox{id}\otimes_{\star}\varepsilon_s^{(e)})\circ\Delta_{\star}=(\varepsilon_s^{(e')}\otimes_{\star}\mbox{id})
\circ\Delta_{\star}=\mbox{id},\quad\forall e,e'\in G^{(0)}$,\item $4_3.\ \
(\mbox{id}\otimes_{\star}\varepsilon_r^{(e)})\circ\Delta_{\star}=(\varepsilon_s^{(e')}\otimes_{\star}\mbox{id})
\circ\Delta_{\star}=\mbox{id},\quad\forall e,e'\in G^{(0)}$,\item $4_4.\ \
(\mbox{id}\otimes_{\star}\varepsilon_s^{(e)})\circ\Delta_{\star}=(\varepsilon_r^{(e')}\otimes_{\star}\mbox{id})
\circ\Delta_{\star}=\mbox{id},\quad\forall e,e'\in G^{(0)}$,\end{description}with a set of compatibility
conditions chosen among the following (or a suitable combination of them)
\begin{description}\item $5_1.\ \ \varepsilon_r^{(e)}\circ\eta_r^{(e')}=\varepsilon_r^{(e)}\circ\eta_r^{(e)}=
\mbox{id},\ \ \eta_r^{(e)}\circ\varepsilon_r^{(e')}=\eta_r^{(e)}\circ\varepsilon_r^{(e')}=\mbox{id}_{G^{(0)}},\
\ \forall e,e'\in G^{(0)}$,\item $5_2.\ \
\varepsilon_s^{(e)}\circ\eta_s^{(e')}=\varepsilon_s^{(e)}\circ\eta_s^{(e)}= \mbox{id},\ \
\eta_s^{(e)}\circ\varepsilon_s^{(e')}=\eta_s^{(e)}\circ\varepsilon_s^{(e')}=\mbox{id}_{G^{(0)}},\ \ \forall
e,e'\in G^{(0)}$,\item $5_3.\ \ \varepsilon_r^{(e)}\circ\eta_s^{(e')}=\varepsilon_r^{(e)}\circ\eta_s^{(e)}=
\mbox{id},\ \ \eta_r^{(e)}\circ\varepsilon_s^{(e')}=\eta_r^{(e)}\circ\varepsilon_s^{(e')}=\mbox{id}_{G^{(0)}},\
\ \forall e,e'\in G^{(0)}$,\item $5_4.\ \
\varepsilon_s^{(e)}\circ\eta_r^{(e')}=\varepsilon_s^{(e)}\circ\eta_r^{(e)}= \mbox{id},\ \
\eta_s^{(e)}\circ\varepsilon_r^{(e')}=\eta_s^{(e)}\circ\varepsilon_r^{(e')}=\mbox{id}_{G^{(0)}},\ \ \forall
e,e'\in G^{(0)};$\end{description}in such a case, we may define a suitable \it linear $\mathbb{K}$-coalgebroid
\rm structure of the type $(\mathfrak{G}_{\mathbb{K}},\Delta_{\star}, \{\varepsilon_r^{(e)}\}_{e\in G^{(0)}},
\{\varepsilon_s^{(e)}\}_{e\in G^{(0)}})$, whence a \it linear $\mathbb{K}$-coalgebroid \rm structure of the type
$(\mathfrak{G}_{\mathbb{K}},m_{\star},\Delta_{\star},\{\eta_r^{(e)}\}_{e\in G^{(0)}},\{\eta_s^{(e)}\}_{e\in
G^{(0)}}, \{\varepsilon_r^{(e)}\}_{e\in G^{(0)}},\{\varepsilon_s^{(e)}\}_{e\in G^{(0)}})$.\\

If, when it is possible, we impose certain $\mathbb{K}$-algebroid homomorphism conditions for the maps
$\Delta_{\star}, \{\varepsilon_r^{(e)}\}_{e\in G^{(0)}}, \{\varepsilon_s^{(e)}\}_{e\in G^{(0)}}$, and/or certain
$\mathbb{K}$-coalgebroid homomorphism conditions for the maps $m_{\star}, \{\eta_r^{(e)}\}_{e\in G^{(0)}},
\{\eta_s^{(e)}\}_{e\in G^{(0)}}$, then we may to establish a certain \it linear $\mathbb{K}$-bialgebroid \rm
structure on $\mathfrak{B}_{\mathfrak{G}_{\mathbb{K}}}$, having posed
$\mathfrak{B}_{\mathfrak{G}_{\mathbb{K}}}=(\mathfrak{G}_{\mathbb{K}},m_{\star},\Delta_{\star},\{\eta_r^{(e)}\}_{e\in
G^{(0)}},\{\eta_s^{(e)}\}_{e\in G^{(0)}}, \{\varepsilon_r^{(e)}\}_{e\in G^{(0)}},\{\varepsilon_s^{(e)}\}_{e\in
G^{(0)}})$.\\

Finally, if, for each $f,g\in End\ (G_{\mathbb{K}})$ such that $f\otimes_{\star}g$ there exists, we put
$$G\stackrel{\Delta_{\star}}{\longrightarrow}G\otimes_{\star}G\stackrel{f\otimes_{\star}g}{\longrightarrow}
G\otimes_{\star}G\stackrel{m_{\star}}{\longrightarrow}G,$$then it is possible to consider the following
(partial) convolution product$$f\tilde{\ast}_{\star}g\doteq m_{\star}\circ
(f\otimes_{\star}g)\circ\Delta_{\star}\in End\ (G_{\mathbb{K}}).$$Thus, an element $a\in End\ (G_{\mathbb{K}})$
may to be said an \it antipode \rm of a $\mathbb{K}$-bialgebroid structure
$\mathfrak{B}_{\mathfrak{G}_{\mathbb{K}}}$, when there exists $a\tilde{\ast}_{\star}\mbox{id}$,
$\mbox{id}\tilde{\ast}_{\star}a$, and$$a{\tilde{\ast}}_{\star}\mbox{id}=\mbox{id}\tilde{\ast}_{\star}a=\mbox{\rm
$2^{th}$ condition of}\ 5_i,$$if $\mathfrak{B}_{\mathfrak{G}_{\mathbb{K}}}$ has the property $5_i,\ \
i=1,2,3,4$.\\

A $\mathbb{K}$-bialgebroid structure with, at least, one antipode, is said to have an \it E-Hopf algebroid \rm
structure. If $G^{(0)}=\{e\}$, then we obtain an ordinary Hopf algebra structure.$$\bf 5.\ The\ Majid's\
quantum\ gravity\ model$$

\rm S. Majid, in [16] and [17], have introduced a particular noncommutative and noncocommutative bicrossproduct
Hopf algebra that should be viewed as a toy model of a physical system in which both quantum effects (the
noncommutativity) and gravitational curvature effects (the noncocommutativity) are unified. The Majid
construction, being a noncommutative noncocommutative Hopf algebra, may be viewed as a non-trivial example of
quantum group having an important physical meaning. We'll apply this model to the EBJ
EBB-groupoid algebra (eventually equipped with a riemannian structure).\\

Following [15, Chap. III, § 1], an E-groupoid $(G,G^{(0)},r,s,\star)$ is said to be a \it differentiable
E-groupoid \rm(or a \it E-groupoid manifold\rm) if $G,G^{(0)}$ are differentiable manifolds, the maps
$r,s:G\rightarrow G^{(0)}$ are surjective submersions, the inclusion $\alpha:G^{(0)}\hookrightarrow G$ is
smooth, and the partial multiplication
$\star:G^{(2)}\rightarrow G$ is smooth (if we understand $G^{(2)}$ as submanifold of $G\times G$).\\
A locally trivial\footnote{For the notion of local triviality of a topological groupoid, see [15, Chap. II, §
2].} differentiable E-groupoid is said to be a \it Lie E-groupoid.\\\rm

Following [10], a differentiable E-groupoid $(G,G^{(0)},r,s,\star)$ is said to be a \it Riemannian E-groupoid
\rm if there exists a metric $g$ over $G$ and a metric $g_0$ over $G^{(0)}$ in such a way that the inversion map
$i:G\rightarrow G$ is an isometry, and $r,s$ are Riemannian submersions of $(G,g_0)$ onto $(G^{(0)},g_0)$.\\

Let $\mathcal{A}_{HBJ}(=\mathcal{A}_{\mathbb{K}}(\mathcal{G}_{HBJ}(\mathcal{F}_I))\cong\mathcal{A}_{\mathbb{K}}
(\mathcal{G}_{Br}(I)))$ be the HBJ EBB-algebra of [11, § 6]: as seen there, it represent the algebra of physical
observables according to the Matrix Quantum Mechanics.\\

The question of which metric to choice, for instance over $\mathcal{A}_{HBJ}$, or over
$\mathcal{G}_{HBJ}(=\mathcal{G}_{HBJ}(\mathcal{F}_I)\cong\mathcal{G}_{Br}(I))$ (for this last groupoid
isomorphism, see [11, § 5]), is not trivial and not a priori dictated (see, [12, II] for a discussion of a
similar question related to a tentative of metrization of the symplectic phase-space manifold of a dynamical
system, in order that be possible to define a classical Brownian motion on it).\\

Let us introduce a Majid's toy model of quantum mechanics combined with gravity, following [16, 17, 18].\\S.
Majid ([16]) follows the abstract quantization formulation of I. Segal, whereby any abstract
$\mathbb{C}^{\ast}$-algebra can be considered as the algebra of observables of a quantum system, and the
positive linear functionals on it as the states.\\ He, first, consider a pure algebraic formulation of the
classical mechanics of geodesic motion on a Riemannian spacetime manifold, precisely on a homogeneous spacetime,
following the well-known \it Mackey's quantization procedure \rm on homogeneous spacetimes (see [18, Chap. 6],
and references therein).\\

The basis of the Majid's physical picture lies in a new interpretation of the semidirect product algebra as
quantum mechanics on homogeneous spacetime (according to [8]). Namely, he consider the semidirect product
$\mathbb{K}[G_1]\ltimes_{\alpha}\mathbb{K}(G_2)$ where $G_1$  is a finite group that acts, through $\alpha$, on
a set $G_2$; here, $\mathbb{K}[G_1]$ denotes the group algebra over $G_1$, whereas $\mathbb{K}(G_2)$ denotes the
algebra of $\mathbb{K}$-valued functions on $G_2$.\\

[17, section 1.1] motivates the search for self-dual algebraic structures in general, and Hopf algebras in
particular, so that it is natural to search the self-dual structure of
$\mathbb{K}[G_1]\ltimes_{\alpha}\mathbb{K}(G_2)$, as follows.\\

To this end, we assume that $G_2$ is also a group that acts back by an action $\beta$ on $G_1$ as a set; so, one
can equivalently view that $\beta$ induces a coaction of $\mathbb{K}[G_2]$ on $\mathbb{K}(G_1)$, and defines the
corresponding semidirect coproduct coalgebra which we denote $\mathbb{K}[G_1]^{\beta}\rtimes\mathbb{K}(G_2)$.
Such a bicrossproduct structure will be denoted
$\mathbb{K}[G_1]^{\beta}\bowtie_{\alpha}\mathbb{K}(G_2)$.\\

Majid's model fit together the semidirect product by $\alpha$ with the semidirect coproduct by $\beta$, to form
a Hopf algebra; in such a way, we'll have certain (compatibility) constraints on $(\alpha,\beta)$ that gives a
\it bicrossproduct \rm Hopf algebra structure to $\mathbb{K}[G_1]\otimes\mathbb{K}(G_2)$, that it is of
self-dual type; this structure is non-commutative [non-cocommutative] when $\alpha$ [$\beta$] is non-trivial.\\

Majid's model of quantum gravity starts from the physical meaning of a bicrossproduct structure relative to the
case $\mathbb{K}=\mathbb{C}$, $G_1=G_2=\mathbb{R}$ and $\alpha_{\rm left\ u}(s)=\hbar u+s$, with $\hbar$ a
dimensionful parameter (Planck's constant), achieved as a particular case of the classical self-dual $\ast$-Hopf
algebra of observables (according to I. Segal) $\mathbb{C}^{\ast}(G_1)\otimes\mathbb{C}(G_2)$, where $G_1,G_2$
has a some group structure and $\mathbb{C}^{\ast}(G_1)$ is the convolution $\mathbb{C}^{\ast}$-algebra on $G_1$.
Further, with a suitable compatibility conditions (see [16, (9)] or [18, (6.15)]) for $(\alpha,\beta)$ $-$ that
can be viewed as certain (Einstein) ''second-order gravitational field equations'' for $\alpha$ (that induces
metric properties on $G_2$), with back-reaction $\beta$ playing the role of an auxiliary physical field $-$ we
obtain a bicrossproduct Hopf algebra $\mathbb{C}^{\ast}(G_1)^{\beta}\bowtie_{\alpha}\mathbb{C}(G_2)$, with the
following self-duality
$(\mathbb{C}^{\ast}(G_1)^{\beta}\bowtie_{\alpha}\mathbb{C}(G_2))^{\ast}\cong\mathbb{C}^{\ast}(G_2)^{\alpha}
\bowtie_{\beta}\mathbb{C}(G_1)$.\\Moreover (see [16]), in the Lie group setting, the non-commutativity of $G_2$
(whence, the non-cocommutativity of the coalgebra structure) means that the intrinsic torsion-free connection on
$G_2$, has curvature (cogravity), that is to say, the non-cocommutativity plays the role of a Riemannian
curvature on $G_2$ (in the sense of non-commutative geometry).\\

We, now, consider a simple quantization problem (see [17, § 1.1.2]). Let $G_1=G_2$ be a group and $\alpha$ the
left action; hence, the algebra $\mathcal{W}(G)\doteq\mathbb{K}^{\ast}[G]\ltimes_{\rm left}\mathbb{K}(G)$ will
be called the \it group Weyl algebra \rm of $G$, and it represents the algebraic quantization of a particle
moving on $G$ by translations.\\

Finally, if we want to apply these considerations to the case $G_1=G_2=\mathcal{G}_{HBJ}$, then we must consider
both the finite-dimensional and the infinite-dimensional case, in such a way that be possible to determine the
dual (or the restricted dual) of $\mathcal{A}_{\mathbb{K}}(\mathcal{G}_{HBJ})$.\\

Taking into account the physical meaning of $\mathcal{G}_{HBJ}$, it follows that it is possible to consider the
above mentioned algebraic structures (with their physical meaning) in relation to the case study\footnote{Such a
groupoid may be, eventually, endowed with a further topological and/or metric (as the Riemannian one)
structure.} $G_1=G_2=\mathcal{G}_{HBJ}$, with consequent physical interpretation (where possible), in such a way
to get non-trivial examples of quantum groupoids\footnote{If we consider a non-commutative non-cocommutative
Hopf algebra as a model of quantum group.} having a possible quantic meaning.$$\bf 6.\ A\ particular\ Weyl\
Algebra\ (and\ other\ structures)$$

The cross and bicross (or double cross) constructions provides the basic algebraic structures on which to build
up non-trivial examples of quantum groups, even in the infinite-dimensional case.\\In this paragraph, we
expose some examples of such constructions.\\

Let $\mathcal{F}_{HBJ}\doteq\mathcal{F}_{\mathbb{K}}(\mathcal{G}_{HBJ}(\mathcal{F}_I))$ be the linear
$\mathbb{K}$-algebra of $\mathbb{K}$-valued functions defined on $\mathcal{G}_{HBJ}$.\\

Let $(G,\cdot)$ be a finite group. If $\mathcal{F}_{\mathbb{K}}(G)$ is the linear $\mathbb{K}$-algebra of
$\mathbb{K}$-valued functions on $G$, then, since
$\mathcal{F}_{\mathbb{K}}(G)\otimes\mathcal{F}_{\mathbb{K}}(G)\cong\mathcal{F}_{\mathbb{K}}(G\times G)$, it
follows that such an algebra can be endowed with a natural structure of Hopf algebra by the following
data\begin{description} \item 1. coproduct
$\Delta:\mathcal{F}_{\mathbb{K}}(G)\rightarrow\mathcal{F}_{\mathbb{K}}(G\times G)$ given by
$\Delta(f)(g_1,g_2)=f(g_1\cdot g_2)$ for all $g_1,g_2\in G$;\item 2. counit
$\varepsilon:\mathcal{F}_{\mathbb{K}}(G)\rightarrow\mathbb{K}$, with $\varepsilon(f)=1$ for each $f\in G$;\item
3. antipode $S:\mathcal{F}_{\mathbb{K}}(G)\rightarrow\mathcal{F}_{\mathbb{K}}(G)$, defined by
$S(f)(g)=f(g^{-1})$ for all $g\in G$.\end{description}If we consider a finite groupoid instead of a finite
group, then 3. even subsists, but 1. and 2. are no longer valid because of the groupoid structure.\\

If $\mathcal{G}=(G,G^{(0)},r,s,\star)$ is a finite groupoid, then, following [24, § 2.2], the most natural
coalgebra structure on $\mathcal{F}_{\mathbb{K}}(\mathcal{G})$, is given by the following
data\begin{description}\item 1'. coproduct $\Delta(g_1,g_2)=f(g_1\star g_2)$ if $(g_1,g_2)\in G^{(2)}$, and $=0$
otherwise;\item 2'. counit $\varepsilon(f)=\sum_{e\in G^{(0)}}f(e)$.\end{description} In such a way, via 1', 2'
and 3, it is possible to consider a Hopf algebra structure on $\mathcal{F}_{\mathbb{K}}(\mathcal{G})$.\\

In the finite-dimensional case, we have
$\mathcal{A}^{\ast}_{\mathbb{K}}(\mathcal{G})\cong\mathcal{F}_{\mathbb{K}}(\mathcal{G})$ (see [18, Example
1.5.4]) as regard the dual Hopf algebra $\mathcal{A}^{\ast}_{\mathbb{K}}(\mathcal{G})$ of
$\mathcal{A}_{\mathbb{K}}(\mathcal{G})$, whereas, in the infinite-dimensional case, we have the restricted dual
$\mathcal{A}^{\ast}_{\mathbb{K}}(\mathcal{G})\cong\mathcal{F}^{(o)}_{\mathbb{K}}(\mathcal{G})
\subseteq\mathcal{F}_{\mathbb{K}}(\mathcal{G})$; hence, in our case $\mathcal{G}=\mathcal{G}_{HBJ}$, following
[18, Chap. 6] it is possible to consider a non-degenerate dual pairing, say $\langle\ ,\ \rangle$, between
$\mathcal{A}_{\mathbb{K}}(\mathcal{G}_{HBJ})$ and $\mathcal{F}^{(o)}_{\mathbb{K}}(\mathcal{G}_{HBJ})$ (in the
finite-dimensional case, it is
$\mathcal{F}^{(o)}_{\mathbb{K}}(\mathcal{G}_{HBJ})=\mathcal{F}_{\mathbb{K}}(\mathcal{G}_{HBJ})$).\\
Hence, if we define the action$$\alpha:(b,a)\rightarrow b\rhd a=\langle b,a_{(1)}\rangle a_{(2)}\qquad \forall
a\in \mathcal{A}_{\mathbb{K}}(\mathcal{G}_{HBJ}),\ \forall
b\in\mathcal{F}^{(o)}_{\mathbb{K}}(\mathcal{G}_{HBJ}),$$ then it is possible to define the left cross product
algebra
$$\mathcal{H}(\mathcal{A}_{\mathbb{K}}(\mathcal{G}_{HBJ})\doteq\mathcal{A}_{\mathbb{K}}(\mathcal{G}_{HBJ})
\ltimes_{\alpha}\mathcal{F}^{(o)}_{\mathbb{K}}(\mathcal{G}_{HBJ}),$$called the \it Heisenberg double \rm of
$\mathcal{A}_{\mathbb{K}}(\mathcal{G}_{HBJ})$.\\

If $V$ is an $A$-module algebra, with $A$ a Hopf algebra, then let $V\ltimes A$ be the corresponding left cross
product, and $(v\otimes a)\rhd w=v(a\rhd w)$ the corresponding Schr\"{o}dinger representation (see [18, § 1.6])
of $V$ on itself.\\If $V$ has a Hopf algebra structure and $V^{(o)}$ is its restricted dual via the pairing
$\langle\ ,\ \rangle$, then (see [18, Chap. 6]) the action $\sigma$ given by $\phi\rhd
v=v_{(1)}\langle\phi,v_{(2)}\rangle$ for all $v\in V,\phi\in V^{(o)}$, make $V$ into a $V^{(o)}$-module algebra
and $V\otimes V^{(o)}$ into an algebra with product given by
$$(v\otimes\phi)(w\otimes\psi)=vw_{(1)}\otimes\langle w_{(2)},\phi_{(1)}\rangle\phi_{(2)}\psi,$$so that let
$\mathcal{W}(V)\doteq V\ltimes_{\sigma} V^{(o)}$ be the corresponding left cross product algebra. Hence, it is
possible to prove (see [18, Chap. 6]) that the related Schr\"{o}dinger representation (see [18, § 6.1, p. 222])
give rise to an algebra isomorphism $\chi:V\ltimes_{\sigma}V^{(o)}\rightarrow Lin(V)$ (= algebra of
$\mathbb{K}$-endomorphisms of $V$), given by $\chi(v\otimes\psi)w=vw_{(1)}\langle\phi,w_{(2)}\rangle$;
$\mathcal{W}(V)\doteq V\ltimes_{\sigma} V^{(o)}$ is said to be the (\it restricted\rm) \it group Weyl algebra
\rm of the Hopf algebra $V$.\\

This last construction is an algebraic generalization of the usual Weyl algebra of Quantum Mechanics on a group
(see [17, § 1.1.2]), whose finite-dimensional prototype is as follows.

Let $G$ be a finite group, and let's consider the strict dual pair given by the $\mathbb{K}$-valued functions on
$G$, say $\mathbb{K}(G)$, and the free algebra on $G$, say $\mathbb{K}G$. Hence, the right action of $G$ on
itself given by $\psi_u(s)=su$, establishes a left cross product algebra structure, say
$\mathbb{K}(G)\ltimes\mathbb{K}G$, on $\mathbb{K}(G)\otimes\mathbb{K}G$; such an action induces (see [18, Chap.
6]), also, a left regular representation of $G$ into $\mathbb{K}G$, so that we can consider the related
Schr\"{o}dinger representation generated by it and by the action of $\mathbb{K}G$ on itself by pointwise
product. Thus, if $V=\mathbb{K}(G)$, with $\mathbb{K}(G)$ endowed with the usual Hopf algebra structure, then we
have that the Weyl algebra $\mathbb{K}(G)\ltimes\mathbb{K}G$ (with $V^{(o)}=\mathbb{K}(G)^{\ast}=\mathbb{K}G$
since $G$ is finite) is isomorphic to $Lin(\mathbb{K}(G))$ via the Schr\"{o}dinger representation. As already
said in the previous paragraph, such a Weyl algebra formalizes the algebraic quantization of a particle moving
on $G$ by
translations.\\

If we apply what has been said above to $\mathcal{G}_{HBJ}$ in the finite-dimensional case (that is, when \it
card $I<\infty$ \rm that correspond to a finite number of energy levels $-$ see [11, § 6]), since
$V^{(o)}=\mathcal{F}^{(o)}_{\mathbb{K}}(\mathcal{G}_{HBJ})=\mathcal{A}_{\mathbb{K}}(\mathcal{G}_{HBJ})$, then we
have that
$$\mathcal{W}(\mathcal{F}_{\mathbb{K}}(\mathcal{G}_{HBJ}))=\mathcal{F}_{\mathbb{K}}(\mathcal{G}_{HBJ})\ltimes
\mathcal{A}_{\mathbb{K}}(\mathcal{G}_{HBJ})$$represents the algebraic quantization of a particle moving on the
groupoid $\mathcal{G}_{HBJ}$ by translations\footnote{From here, we may speak of a Quantum Mechanics on a
groupoid.}, remembering the quantic meaning (according to I. Segal) of
$\mathcal{F}_{\mathbb{K}}(\mathcal{G}_{HBJ})$ (as set of states) and
$\mathcal{A}_{\mathbb{K}}(\mathcal{G}_{HBJ})$ (as set of observables).\\Instead, in the infinite-dimensional
case, we have that $\mathcal{F}^{(o)}_{\mathbb{K}}(\mathcal{G}_{HBJ})$ is isomorphic to a sub-Hopf algebra of
$\mathcal{A}_{\mathbb{K}}(\mathcal{G}_{HBJ})$ (this is the HBJ EBBH-algebra $-$ see [11, § 8]), so that
$$\mathcal{W}(\mathcal{F}_{\mathbb{K}}(\mathcal{G}_{HBJ}))=\mathcal{F}_{\mathbb{K}}(\mathcal{G}_{HBJ})\ltimes
\mathcal{F}^{(o)}_{\mathbb{K}}(\mathcal{G}_{HBJ})\subseteq\mathcal{F}_{\mathbb{K}}(\mathcal{G}_{HBJ})\ltimes
\mathcal{A}_{\mathbb{K}}(\mathcal{G}_{HBJ}).$$Finally, the right adjoint action (see [18, § 1.6]) of
$\mathcal{G}_{HBJ}$ on itself given by $\psi_g(h)=g^{-1}\star h\star g$ if there exists, and $=0$ otherwise,
make $\mathcal{F}_{\mathbb{K}}(\mathcal{G}_{HBJ})$ into an $\mathcal{A}_{\mathbb{K}}(\mathcal{G}_{HBJ})$-module
algebra. In the finite-dimensional case, we have $\mathcal{A}^{\ast}_{\mathbb{K}}(\mathcal{G}_{HBJ})\cong
\mathcal{F}_{\mathbb{K}}(\mathcal{G}_{HBJ})$, so that
$$\mathcal{A}_{\mathbb{K}}(\mathcal{G}_{HBJ})=\mathcal{A}^{\ast\ast}_{\mathbb{K}}(\mathcal{G}_{HBJ})\cong
\mathcal{F}^{\ast}_{\mathbb{K}}(\mathcal{G}_{HBJ}),$$whence $\mathcal{F}_{\mathbb{K}}(\mathcal{G}_{HBJ})$ is a
$\mathcal{F}^{\ast}_{\mathbb{K}}(\mathcal{G}_{HBJ})$-module algebra too. Therefore, we may consider the
following left cross product algebra
$$\mathcal{F}_{\mathbb{K}}(\mathcal{G}_{HBJ})\ltimes\mathcal{F}^{\ast}_{\mathbb{K}}(\mathcal{G}_{HBJ})\cong
\mathcal{F}_{\mathbb{K}}(\mathcal{G}_{HBJ})\ltimes\mathcal{A}_{\mathbb{K}}(\mathcal{G}_{HBJ})$$that the tensor
product coalgebra makes into a Hopf algebra, called the (\it Drinfeld\rm) \it quantum double \rm of
$\mathcal{G}_{HBJ}$, and denoted $\mathcal{D}(\mathcal{G}_{HBJ})$; even in the finite-dimensional case, it
represent the algebraic quantization of a particle constrained to move on conjugacy classes of
$\mathcal{G}_{HBJ}$ (quantization on homogeneous space over a groupoid).\\

Besides, it has been proved, for a finite group $G$, that this (Drinfeld) quantum double
$\mathcal{D}(\mathcal{G}_{HBJ})$, has a quasitriangular structure (see [18, Chap. 6]) given by
$$(\delta_s\otimes u)(\delta_t\otimes v)=\delta_{u^{-1}su,t}\delta_t\otimes uv,\qquad \Delta(\delta_s\otimes u)=
\sum_{ab=s}\delta_a\otimes u\delta_b\otimes u,$$ $$\varepsilon(\delta_s\otimes u)=\delta_{s,e},\qquad
S(\delta_s\otimes u)=\delta_{u^{-1}s^{-1}u}\otimes u^{-1},$$ $$R=\sum_{u\in G}\delta_u\otimes e\otimes 1\otimes
u,$$where we have identifies the dual of $\mathbb{K}G$ with $\mathbb{K}(G)$ via the idempotents $p_g,g\in G$
such that $p_gp_h=\delta_{g,h}p_g$ (see [19, § 2.5] and [18, § 1.5.4]).\\ Such a quantum double represents the
algebra of quantum observables of a certain physical system with symmetry group $G$.\\

Hence, even in the finite-dimensional case\footnote{The infinite-dimensional case is not so immediate.}, we may
consider, with suitable modifications, an analogous quasitriangular structure on $\mathcal{G}_{HBJ}$, obtaining
the (Drinfeld) quantum double $\mathcal{D}(\mathcal{G}_{HBJ})$ on $\mathcal{G}_{HBJ}$; thus, if we consider a
quasitriangular Hopf algebra as a model of quantum group, the (Drinfeld) quantum double
$\mathcal{D}(\mathcal{G}_{HBJ})$ provides a non-trivial example of quantum group having a (possible) quantic
meaning (related to a quantum mechanics on a groupoid).$$\bf 7.\ Conclusions$$

From what has been said above, and in [11], it rises a possible role played by the groupoid structures in
Quantum Mechanics and Quantum Field
Theory.\\

For instance, such a groupoid structures\footnote{Eventually equipped with further, more specific structures, as
the topological or metric ones.} might takes place a prominent rule in Renormalization, as well as in the
Majid's model of quantum gravity. Indeed, a central problem in quantum gravity concerns its nonrenormalizability
due to the existence of UV divergences; in turn, the UV divergences arise from the assumption that the classical
configurations being summed over are defined on a continuum.\\

So, the discreteness given by groupoid structures may turn out to be of some usefulness in such a
(renormalization) QFT problems.

\end{document}